\def\bsg{$b \rightarrow s \gamma$}
\def\bsgeq{b \rightarrow s \gamma}
\def\Bbsg{B(\bsgeq)}
\def\ttt{\theta_{\tilde t}}
\def\mst1{m_{\tilde t_1}}
\def\msqk{\tilde m}
\def\sign{\mbox{\rm sign}}
\def\etal{{\it et. al.}}

\def\Ref{\bibitem}

\def\noi{\noindent}
\def\pref#1{(\ref{#1})}
\def\beq{\begin{equation}}
\def\eeq{\end{equation}}
\def\bmulteq{\begin{eqnarray}}
\def\emulteq{\end{eqnarray}}

\def\SM{Standard Model}

\def\susyic{supersymmetric}

\def\abs #1{|#1|}

\documentstyle[12pt]{article}

\begin{document}
%\draft

\def\preprint#1{{\vskip -.25in\hfill\hfil#1\vskip 10pt}}

\vskip -1 in
\preprint{TRI-PP-93-66}
\preprint{July 1993}

% title
\centerline{{\Large\bf
 Supersymmetric $b \rightarrow s \gamma$ with Large Chargino Contributions }}
\centerline{R. Garisto and J. N. Ng}
%
%\begin{instit}
\begin{center}
{\it TRIUMF,
4004 Wesbrook Mall,
Vancouver, B.C.,
V6T 2A3,
Canada}
\end{center}
%\end{instit}

%
\begin{abstract}
Supersymmetric (SUSY) theories are often thought to give large branching ratios
for \bsg\ from charged Higgs loops.  We show that in many cases
chargino loop contributions can cancel those of the Higgs, and SUSY can
give $\Bbsg$ at or below the \SM\ prediction.  We show this occurs because
the large stop mass splittings usually found in
SUSY break a GIM mechanism suppression.  These effects are strongly enhanced
by large $\tan\beta$, so that $\Bbsg$ is very sensitive to the value
of $\tan\beta$, contrary to what has been claimed.
We also note that the supergravity relation $B_0 = A_0-1$ is somewhat
disfavored over the general case.
\end{abstract}

%%%%%%%%%%%%%%%%%%%%%%%%%%%%%%%%%%%%%%%%%%%%%%%%%%%%%%%%%%%%%%%%%%%%%%%%%%

%% Intro

There has been much interest in the decay \bsg\ because of new results
from the CLEO collaboration which bound the inclusive branching ratio,
$\Bbsg$, below
$5.4\times 10^{-4}$ at the $95\%$ confidence level, and give a non-zero
branching ratio for the exclusive decay $B\rightarrow K^* \gamma$ of about
$5\times 10^{-5} $\cite{CLEO new}. One expects this exclusive
channel to make up $5\%-40\%$ of the inclusive rate \cite{exclusive},
so $\Bbsg$ must be {\it greater} than about $10^{-4}$.
The \SM\ (SM) contribution depends slowly on the top quark mass and is
of order $4\times 10^{-4}$ for $m_t$ of 140 GeV.  Given this, some recent works
\cite{Barger H,Hewett H} claim that the charged Higgs ($H^+$)
masses in supersymmetric theories \cite{HaK}
must be very large to avoid exceeding the upper bound on $\Bbsg$.
We show that this is not always the case---that chargino
($\chi^+$) loop contributions can
cancel the $H^+$ contributions and give $\Bbsg$ near or below the SM
prediction. In particular, we show that such destructive interference
effects are important for large $\tan\beta$
(which is the ratio of Higgs vacuum expectation values),
and when there is a large
stop mass splitting.
We show that the latter effect is due to the breaking of a GIM
cancellation.

The sign of certain mixing angles is also important because one
needs the chargino contribution to interfere destructively,
rather than constructively.
{}From this one can obtain an approximate condition on the soft SUSY
breaking parameters $A$ and $B$, which may have implications for
supergravity theories (for a review see \cite{Nilles}).

Calculations for $\Bbsg$ in SUSY can be found in the literature
\cite{Bertolini etal,BarbieriG}.
Bertolini \etal\ \cite{Bertolini etal} perform a thorough but
very constrained analysis which imposes radiative breaking,
in the minimal model, with $B_0=A_0-1$.  They also do not consider large
$\tan\beta$, where chargino effects can become much more important.
Barbieri and Giudice \cite{BarbieriG} make the important point that
$\Bbsg$ vanishes in the exact \susyic\ limit.
However, the scenarios they consider (which are
indeed close to the SUSY limit) with gaugino mass ($m_\lambda$) and Higgs
mixing
mass ($\mu$) set to zero, are not phenomenologically viable
because they give chargino and neutralino masses which are too small (one
of the higgsinos is even massless in this case).
These approaches
are understandable since there are many parameters in SUSY theories.
Our approach is to concentrate on those parameters which tend to make the
chargino contribution large and destructively interfering, so as to make
qualitative statements about what areas of parameter space are favored.
We show that one cannot neglect the chargino contributions and that
there are large areas of parameter space where $\Bbsg$ in SUSY is at or below
the SM prediction.  Contrary to what is claimed in \cite{BarbieriG},
we find that $\Bbsg$ is very sensitive to $\tan\beta$, and one can even find
regions for large $\tan\beta$ where the chargino destructive interference
is too large \cite{foot why wrong}.

%BSG

The inclusive decay \bsg\ comes from the operator
$\bar s_L \sigma^{\mu\nu} b_R F_{\mu\nu}$.
When one runs the scale from $M_Z$ to $m_b$, this operator mixes with
the gluon operator
$\bar s_L \sigma^{\mu\nu} T^a b_R G_{\mu\nu}^a$,
as well as four quark operators.
We use the notation of \cite{BarbieriG} throughout, up to an overall
sign in the amplitude.
They define the coefficients of the photon (and gluon) operators as
$G_F (\alpha /8\pi^3)^{1/2} V_{ts}^* V_{tb} m_b\, A_\gamma$
(and $A_\gamma \rightarrow A_g$).
We will concentrate on the photon coefficient $A_\gamma$ because the gluon
coefficient contribution is relatively suppressed by
QCD factors \cite{QCD impt}, as can be seen from the ratio of
inclusive branching ratios \cite{BarbieriG}:
\beq
{B(\bsgeq) \over B(b \rightarrow ce\nu)} = {6 \alpha \over \pi}\,
{ \left| \eta^{16 \over 23} A_\gamma + {8\over 3}
(\eta^{14\over 23} - \eta^{16 \over 23}) A_g + C \right|^2  \over
I(m_c/m_b) \left( 1 - {2 \over 3\pi}\alpha_s(m_b) f(m_c/m_b) \right) },
\eeq

\noi
where the inclusive semileptonic branching ratio is
$B(b \rightarrow ce\nu) \simeq 0.107$; the QCD factor
$\eta = \alpha_s(M_Z)/\alpha_s(m_b) \simeq 0.546$; a QCD correction factor
$f(m_c/m_b)\simeq 2.41$; and $I(x) = 1-8x^2+8x^6-x^8-24x^4\ln x$ is
a phase space factor. The constant $C$ comes from mixing of four
quark operators as we run down to $m_b$, and is about $0.175$ \cite{BarbieriG}.

The photon operator coefficient $A_\gamma$ comes mainly
from loops with a $W^+$ and a top quark,
an $H^+$ and a top quark, and charginos $\chi_j^+$ (j=1,2) and
up squarks.  There are also contributions from flavor changing neutral current
(FCNC) vertices
due to squark flavor mixing, but these contributions
tend to be very small in the minimal model \cite{Bertolini etal}.
Squark flavor mixings can be large in certain non-minimal models, but
they are constrained to be small by other FCNC
observables so that their contribution to \bsg\ is generally small
\cite{Hagelin}.
Thus we can write

\beq
A_\gamma \simeq A_\gamma(W^+) + A_\gamma(H^+) + A_\gamma(\chi^+),
\eeq

\noi
where $A_\gamma(W^+)$ and $A_\gamma(H^+)$ are always greater than zero,
while $A_\gamma(\chi^+)$ can be of either sign.
In the limit of degenerate up and charm squark masses,
we can write
$A_\gamma(\chi^+) = A_1(\chi^+) + A_2(\chi^+) + A_3(\chi^+) + A_4(\chi^+)$,
with \cite{BarbieriG}
\bmulteq
&&A_1(\chi^+) \simeq +\sum_{j=1}^2 {m_W^2 \over \tilde m_{\chi_j}^2}
|V_{j1}|^2\, g^{(1)}({x_{0}}_j),\\
&&A_2(\chi^+) \simeq -\sum_{j,k=1}^2 {m_W^2 \over \tilde m_{\chi_j}^2}
\left|V_{j1} T_{k1} - V_{j2} T_{k2} {m_t \over v_u}
 \right|^2 \, g^{(1)}({x_{k}}_j),\\
&&A_3(\chi^+) \simeq - \sum_{j=1}^2 {m_W \over \tilde m_{\chi_j}}
{ U_{j2} V_{j1} \over \sqrt{2} \cos\beta} \, g^{(3)}({x_{0}}_j),\\
&&A_4(\chi^+) \simeq +\sum_{j,k=1}^2 {m_W \over \tilde m_{\chi_j}}
{ U_{j2} \left( V_{j1} {T_{k1}}^2 - V_{j2} T_{k1} T_{k2} {m_t \over v_u}
\right)
  \over \sqrt{2} \cos\beta} \, g^{(3)}({x_{k}}_j),
\emulteq

\noi
where $U_{ij}$ and $V_{ij}$ are the unitary matrices which diagonalize
the chargino mass matrix (see \cite{HaK}),
$T_{kl}$ diagonalizes the stop mass matrix,
%% \! is to avoid overfull hbox %%
$v_u \!=\! \sqrt{2} m_W \sin\beta$, and we define
${x_0}_j \equiv \msqk^2/\tilde m_{\chi_j}^2$ and
${x_k}_j \equiv \msqk_{t_k}^2/\tilde m_{\chi_j}^2$.
We have used

\beq
g(\msqk_u^2/\msqk_{\chi_j}^2) V_{us}^* V_{ub} +
g(\msqk_c^2/\msqk_{\chi_j}^2) V_{cs}^* V_{cb} \simeq
-g(\msqk^2/\msqk_{\chi_j}^2) V_{ts}^* V_{tb}
\label{GIM eq}
\eeq

\noi
which follows from the unitarity condition $V_{\alpha s}^* V_{\alpha b}=0$ and
the condition that the first two generations of up squarks are nearly
degenerate.
The functions $g(x)$ are given by
\cite{Bertolini etal,BarbieriG}:

\bmulteq
g^{(1)}(x) &&= {8x^3 - 3x^2 - 12x + 7 + (12x - 18x^2)\ln{x} \over 36(x-1)^4},
\\
%%g_2(x) = {5x^2 - 8x + 3 + (4 - 6x)\ln{x} \over 6(x-1)^3},
%%\\
g^{(3)}(x) &&= {-7x^2 + 12x -5 + (6x^2 - 4x)\ln x \over 6 (x-1)^3}.
\emulteq

\noi
Both of these are positive, and fall off as $x$ becomes large.
One finds that $g^{(3)}(x)$ is bigger than $g^{(1)}(x)$
by at least a factor of 4, for all $x$.

We have broken $A_\gamma(\chi)$ up into four pieces to see when it
can significantly reduce $\Bbsg$.  The sum $A_1(\chi)+A_2(\chi)$ is almost
never large enough to cancel the $H^+$ contribution.
On the other hand, $A_3(\chi^+)$
and $A_4(\chi^+)$ can be large because they are enhanced by large $\tan\beta$.
However, if the stop squarks are degenerate in mass with the other up squarks,
these large contributions exactly cancel, due to a GIM cancellation.

% Where BSG is small

To see how the sum $A_3(\chi)+A_4(\chi)$ depends upon the stop mass
splittings, let us define ${f_0}_j \equiv g^{(3)}({x_0}_j)$ and
${f_k}_j \equiv g^{(3)}({x_k}_j) \equiv {f_0}_j + \Delta {f_k}_j$
($j,k = 1,2$).
Defining $\sin\ttt \equiv T_{12}$, we can write

\bmulteq
&&A_3(\chi^+) + A_4(\chi^+)= - \sum_{j=1}^2 {m_W \over \tilde m_{\chi_j}}
{1 \over \sqrt{2} \cos\beta} \nonumber\\
&&\ \ \ \ \ \Bigl[ -U_{j2} V_{j1}
 \left( \cos^2\ttt \Delta {f_1}_j + \sin^2\ttt \Delta {f_2}_j
\right)\nonumber\\
&&\ \ \ \ \ \ + U_{j2} V_{j2}{m_t \over v_u} \sin\ttt \cos\ttt
 \left(\Delta {f_1}_j - \Delta {f_2}_j\right) \Bigr].
\label{A3A4 j}
\emulteq

\noi
One sees immediately that if all the squark masses are degenerate
($\msqk_{t_1} = \msqk_{t_2} = \msqk$), then $\Delta {f_1}_j =\Delta {f_2}_j=0$
which means that $A_3(\chi^+) + A_4(\chi^+)=0$.
{}From \pref{GIM eq} it is clear that this cancellation arises from a
GIM mechanism; if ${x_0}_j = {x_k}_j$, the unitarity of the CKM
matrix ensures that $A_3(\chi^+) + A_4(\chi^+)=0$.

One can get a sense for the behavior of \pref{A3A4 j} by considering
only the light chargino piece ($j=1$), which tends to contribute
more than the heavier chargino since $g^{(3)}(x)$ is larger for
small $x$. A careful analysis of the chargino mass matrix diagonalization
reveals that

\beq
\sign U_{12} V_{11} = - \sign\mu, \ \sign U_{12} V_{12} = + \sign\mu.
\eeq

\noi
The only exception is for $\mu< - \msqk_{wino} \tan\beta$, where
$U_{12} V_{12}$ is positive but very small.  If we use the large
$\tan\beta$ approximation $\cos^{-1}\beta \simeq \tan\beta$, we can
estimate that

\bmulteq
&&A_3(\chi^+) + A_4(\chi^+) \sim - {1 \over \sqrt{2}}
 {m_W \over \tilde m_{\chi_j}} \tan\beta
\nonumber\\
&&\ \ \ \ \ \ \Bigl[ \sign\mu \, \left| U_{12} V_{11} \right|\,
 \left( \cos^2\ttt \, \Delta {f_1} + \sin^2\ttt\, \Delta {f_2} \right)
\nonumber\\
&&\ \ \ \ \ \ \ + \sign\ttt\, \mu \,
\left| U_{12} V_{12} \sin\ttt\, \cos\ttt \right| \,
{m_t \over v_u}  \left(\Delta {f_1} - \Delta {f_2}\right) \Bigr].
\label{A3A4 est}
\emulteq

\noi
This allows one to understand the gross behavior of the sum.
For moderate to large stop splittings, $|\ttt| \sim 45^{^0}$,
$\msqk_{t_1}$ will be less than $\msqk$, and
$\msqk_{t_2}$ will be greater than or of order $\msqk$
\cite{DreesNojiri}.
One sees that the chargino contribution tends to have a large
destructive interference with the $W^+$ and $H^+$ pieces if $\tilde t_1$
is light ($i.e.$ there is a large stop mass splitting),
$\tan\beta$ is large, and if $\ttt\,\mu>0$, $i.e.$

\beq
\mu < 0 , \ttt<0, \ \mbox{\rm or}\ \mu>0, \ttt>0.
\label{mu ttt}
\eeq

\noi
The $\mu>0$ case gives a smaller $\Bbsg$ because both pieces in
\pref{A3A4 est} help to reduce the overall amplitude.
One can show that $\sign\ttt = -\sign(A m_0 - \mu \cot\beta)$, so
that $A \mu<0$ implies that $\ttt\,\mu>0$ (though the converse is not
necessarily true).
Finally we note that the sign of $\mu$ is just the sign of $B$--one
rotates the Higgs fields so as to make the Higgs potential
coefficient $\mu_{12}^2$ positive, and then $\sign\mu$ equals
the $\sign B$ before that rotation \cite{SUSYdn}.
Thus $AB<0$ implies $\ttt\,\mu>0$, which is the favorable region
for destructive interference from the chargino loops.
If $\abs{Am_0}\tan\beta>|\mu|$, the converse is also true.

In the simplest SUGRA theories, one has the relation at the Planck scale
$B_0 = A_0 -1$ \cite{Nilles}. One can show
using general properties of the renormalization group equations
that this relation implies
one {\it cannot} have $A<0$ and $B>0$ at the weak scale,
which is  the most favored region for small
$\Bbsg$. If $m_{H^+}$ and $\tan\beta$
were found experimentally to be small, it might be possible to rule out
minimal SUSY models which satisfy $B_0 = A_0 -1$.

To illustrate these results, we consider some supersymmetric scenarios.
In Figure 1, we consider the heuristic parameters $\Delta \msqk_t$ and
$\ttt$.  We see that for the given choice of parameters with
$\ttt\, \mu<0$, $\Bbsg$ is always greater than
the CLEO bound (in the region allowed by LEP, above the unlabeled curves).
The case $\ttt\, \mu>0$ has lower $\Bbsg$, especially for the $\mu>0$ case,
and there are regions where the CLEO bound is satisfied.
Increasing $\Delta \msqk_t$ lowers
$\Bbsg$ in the $\ttt\, \mu>0$ regions because
$A_3(\chi^+) + A_4(\chi^+)$ becomes more important.
Radiative corrections lower both $\msqk_{t_1}$ and $\msqk_{t_2}$
relative to $\msqk$ \cite{DreesNojiri}, so we take
$(\msqk_{t_k} - \msqk)/\Delta \msqk_t$
to be $-2/3$ and $1/3$, respectively.
If one raises (lowers) $\msqk_{t_2}$ while holding $\msqk_{t_1}$
constant, the difference between the $\mu>0$ and $\mu<0$ regions
tends to become less (more) pronounced.

Figures 2--4 show more realistic scenarios where one inputs $A$
instead of $\Delta \msqk_t$ and $\sign\ttt$.
Increasing $\tan\beta$ will increase $m_{H^+}^2$,
so that large $\tan\beta$ gives smaller $\Bbsg$ just by suppressing the
$H^+$ loop contribution. To examine the different values for $\tan\beta$
on equal footing, we have taken
$\abs{B} = 1.5/\tan\beta$ so that $m_{H^+}^2$ is about the same in each graph
($m_{H^+} \simeq 260$ GeV at $|\mu|=400$).  Even so, $\Bbsg$ gets much smaller
in the $AB<0$ ($i.e.$ $A \mu<0$) regions
as $\tan\beta$ increases,
again because $A_3(\chi^+) + A_4(\chi^+)$ becomes more important.
Larger $|A|$ also reduces
$\Bbsg$ in those regions. For $A<0$, $B>0$ (which is not allowed if
$B_0=A_0-1$) and
$\tan\beta>10$, there are even regions where the chargino contribution
flips the sign of the amplitude---cancelling the $H^+$, $W^+$ and $C$
contributions---so that certain regions of parameter space are ruled
out because the value of $-A_\gamma(\chi)$ is too large!
Conversely, the regions of $AB>0$ tend to give larger $\Bbsg$ due
to constructive interference from $A_\gamma(\chi)$.

We have shown what happens when one varies $\tan\beta$, $A$, $\mu$ and
$m_\lambda$.
We took $|B|$ such that $m_{H^+}^2$ was of order the weak scale---if $|B|$ is
larger (smaller) than in Figures 1--4, $m_{H^+}^2$ will be larger (smaller),
and all the values for $\Bbsg$ will be smaller (larger).
This simply demonstrates the point stressed by \cite{Barger H,Hewett H} that
$\Bbsg$ can be suppressed by large $m_{H^+}$.
If $m_t$ is heavier (lighter) than 140 GeV, all of the values for $\Bbsg$
will be shifted up (down) slightly, but for the
SUSY result of $m_t \simeq 134\pm 25$ GeV \cite{Langacker}, there is no
qualitative change in our results.
Lastly, one can make a different choice for $m_0$. Increasing $m_0$
makes both $\Delta \msqk_t$ and $m_{H^+}^2$ larger,
so that $\Bbsg$ generally decreases in the $AB<0$ regions.
 One must be careful
for large $A m_0$ that $\msqk_{t_1}^2$ is greater than zero.

%%Conclusion

The branching ratio for \bsg\ in SUSY theories is near or below the
SM value if the charged Higgs mass is large, or the chargino contribution
destructively interferes with the charged Higgs and $W$ loops.
We found that the latter
occurs in regions of parameter space where $AB<0$ (or equivalently when
$A \mu <0$), and is accentuated by large
$\tan\beta$ and large $|A|$.
One can have $\Bbsg$ at or below the SM prediction in
\susyic\ models without requiring a large charged Higgs mass.
Finally, we have noted that if $m_{H^+}$ and $\tan\beta$ were found to be
small, it might be possible to rule out minimal SUSY models which satisfy the
SUGRA relation $B_0= A_0 -1$.

\

\centerline{\Large\bf Acknowledgements}

We thank G. Kane and G. Poulis for useful discussions.
This work was supported in part by a grant from the National Sciences
and Engineering Research Council of Canada.

%%%%%%%%%%%%%%%%%%%%%%%%%%%%%%%%%%%%%%%%%%%%%%%%%%%%%%%%%%%%%%%%%%%%%%%%%%%%%%%
%  References
%%%%%%%%%%%%%%%%%%%%%%%%%%%%%%%%%%%%%%%%%%%%%%%%%%%%%%%%%%%%%%%%%%%%%%%%%%%%%%%

\def\vol#1{#1}

%%%%%%%%%%%%%%%%%%%%%%%%%
%% Figure caption
%%%%%%%%%%%%%%%%%%%%%%%%
\newpage

\centerline{\Large\bf Figure Captions}

{\bf Figure 1:} Contour plots of $\Bbsg$ (labeled lines)
in units of $10^{-4}$ for
$\tan\beta=5$, $m_t=140$, $B=0.3$, and $m_0= \abs{\mu}$.
The current CLEO bound in these units is 5.4.
Graphs (a) and (c) ((b) and (d)) use $\ttt <0$ ($\ttt>0$).
Graphs (a) and (b) ((c) and (d)) use a fixed stop mass splitting of
100 (200).  Unlabeled solid lines are $m_{\chi^0_1}=25$,
and $m_{\chi^+_1}=45$. All masses in GeV.

{\bf Figure 2:} Contour plots of $\Bbsg$ (labeled lines)
in units of $10^{-4}$ for
$\tan\beta=3$, $m_t=140$, $B=0.5$, and $m_0= 100$.
Graphs (a), (b), (c), (d) have $A= +1,\ -1,\ +2,\ -2$, respectively.
Unlabeled solid lines are $m_{\chi^0_1}=25$,
and $m_{\chi^+_1}=45$. All masses in GeV.

{\bf Figure 3:} Same as Figure 2 for $\tan\beta=10$.  The contours
in the $\mu>0$ region of (d) curve back near $m_\lambda\sim 80$ because
$A_\gamma(\chi)$ becomes negative enough to flip the sign of the
amplitude.

{\bf Figure 4:} Same as Figure 2 for $\tan\beta=30$.
As in Figure 3, the $\mu>0$ regions of (b) and (d)
have regions which are ruled out because $-A_\gamma(\chi)$ is too large.

\end{document}